\documentclass[cmc,article,final,moreauthors,pdftex]{Definitions/tsp}
\usepackage[normalem]{ulem} 
\usepackage{amsfonts} 
\usepackage{mathcomp} 
\usepackage{CJKutf8} 
\usepackage{pifont} 
\usepackage{bm} 
\usepackage{bbm} 
\graphicspath{{./Definitions/}} 

\makeatletter
\def\T@n@@nc@d@ngM@cr@M@d{}
\def\LY@n@@nc@d@ngM@cr@M@d{}
\makeatother

\let\orignewcommand\newcommand  
\let\newcommand\providecommand  
\usepackage{verse}
\let\newcommand\orignewcommand  

\newsavebox\foobox

\newcommand{\mmathit}[1]{
  \ifthenelse{\equal{#1}{\ln}}{\mathit{ln}}{
    \ifthenelse{\equal{#1}{\max}}{\mathit{max}}{\mathit{#1}}
  }
}
\makeatother
\robustify{\footnote}

\DeclareUnicodeCharacter{1E45}{\.{n}}
\DeclareUnicodeCharacter{1E41}{\.{m}}
\DeclareUnicodeCharacter{2003}{\quad}
\DeclareUnicodeCharacter{2009}{\thinspace}
\DeclareUnicodeCharacter{2002}{\enspace{}}
\DeclareUnicodeCharacter{2005}{\thinspace}
\DeclareUnicodeCharacter{0263}{\textipa{G}}
\DeclareUnicodeCharacter{A0}{~}
\DeclareUnicodeCharacter{2460}{\textcircled{\scriptsize{1}}}
\DeclareUnicodeCharacter{2461}{\textcircled{\scriptsize{2}}}
\DeclareUnicodeCharacter{2462}{\textcircled{\scriptsize{3}}}
\DeclareUnicodeCharacter{2463}{\textcircled{\scriptsize{4}}}
\DeclareUnicodeCharacter{2464}{\textcircled{\scriptsize{5}}}
\DeclareUnicodeCharacter{2465}{\textcircled{\scriptsize{6}}}
\DeclareUnicodeCharacter{2466}{\textcircled{\scriptsize{7}}}
\DeclareUnicodeCharacter{2467}{\textcircled{\scriptsize{8}}}
\DeclareUnicodeCharacter{2468}{\textcircled{\scriptsize{9}}}
\DeclareUnicodeCharacter{2070}{\textsuperscript{0}}
\DeclareUnicodeCharacter{2074}{\textsuperscript{4}}
\DeclareUnicodeCharacter{2075}{\textsuperscript{5}}
\DeclareUnicodeCharacter{2076}{\textsuperscript{6}}
\DeclareUnicodeCharacter{2077}{\textsuperscript{7}}
\DeclareUnicodeCharacter{2078}{\textsuperscript{8}}
\DeclareUnicodeCharacter{2079}{\textsuperscript{9}}
\DeclareUnicodeCharacter{02C2}{<}
\DeclareUnicodeCharacter{2033}{\relax\ifmmode '' \else $''$\fi}
\DeclareUnicodeCharacter{2034}{\relax\ifmmode ''' \else $'''$\fi}
\DeclareUnicodeCharacter{2026}{\relax\ifmmode … \else $\ldots$\fi}
\DeclareUnicodeCharacter{0229}{\c{e}}
\DeclareUnicodeCharacter{016F}{\r{u}}
\DeclareUnicodeCharacter{127}{\relax\ifmmode\rm\hbar\else $\rm\hbar$\fi}
\DeclareUnicodeCharacter{3AC}{\relax\ifmmode\acute{\alpha}\else $\acute{\alpha}$\fi}
\DeclareUnicodeCharacter{3AD}{\relax\ifmmode\acute{\varepsilon}\else $\acute{\varepsilon}$\fi}
\DeclareUnicodeCharacter{3AE}{\relax\ifmmode\acute{\eta}\else $\acute{\eta}$\fi}
\DeclareUnicodeCharacter{3AF}{\relax\ifmmode\acute{\iota}\else $\acute{\iota}$\fi}
\DeclareUnicodeCharacter{3CC}{\relax\ifmmode\acute{o}\else $\acute{o}$\fi}
\DeclareUnicodeCharacter{3CD}{\relax\ifmmode\acute{\upsilon}\else $\acute{\upsilon}$\fi}
\DeclareUnicodeCharacter{3CE}{\relax\ifmmode\acute{\omega}\else $\acute{\omega}$\fi}
\DeclareUnicodeCharacter{391}{A}
\DeclareUnicodeCharacter{392}{B}
\DeclareUnicodeCharacter{395}{E}
\DeclareUnicodeCharacter{396}{Z}
\DeclareUnicodeCharacter{397}{H}
\DeclareUnicodeCharacter{399}{I}
\DeclareUnicodeCharacter{39A}{K}
\DeclareUnicodeCharacter{39C}{M}
\DeclareUnicodeCharacter{39D}{N}
\DeclareUnicodeCharacter{39F}{O}
\DeclareUnicodeCharacter{3A1}{P}
\DeclareUnicodeCharacter{3A4}{T}
\DeclareUnicodeCharacter{3A7}{X}

\DeclareUnicodeCharacter{27E6}{\relax\ifmmode \llbracket \else $\llbracket$\fi}
\DeclareUnicodeCharacter{27E7}{\relax\ifmmode \rrbracket \else $\rrbracket$\fi}
\DeclareUnicodeCharacter{00B5}{\relax\ifmmode \mu{} \else $\mu$\fi}
\DeclareUnicodeCharacter{1D434}{\relax\ifmmode A \else $A$\fi}
\DeclareUnicodeCharacter{1D435}{\relax\ifmmode B \else $B$\fi}
\DeclareUnicodeCharacter{1D436}{\relax\ifmmode C \else $C$\fi}
\DeclareUnicodeCharacter{1D437}{\relax\ifmmode D \else $D$\fi}
\DeclareUnicodeCharacter{1D438}{\relax\ifmmode E \else $E$\fi}
\DeclareUnicodeCharacter{1D439}{\relax\ifmmode F \else $F$\fi}
\DeclareUnicodeCharacter{1D43A}{\relax\ifmmode G \else $G$\fi}
\DeclareUnicodeCharacter{1D43B}{\relax\ifmmode H \else $H$\fi}
\DeclareUnicodeCharacter{1D43C}{\relax\ifmmode I \else $I$\fi}
\DeclareUnicodeCharacter{1D43D}{\relax\ifmmode J \else $J$\fi}
\DeclareUnicodeCharacter{1D43E}{\relax\ifmmode K \else $K$\fi}
\DeclareUnicodeCharacter{1D43F}{\relax\ifmmode L \else $L$\fi}
\DeclareUnicodeCharacter{1D440}{\relax\ifmmode M \else $M$\fi}
\DeclareUnicodeCharacter{1D441}{\relax\ifmmode N \else $N$\fi}
\DeclareUnicodeCharacter{1D442}{\relax\ifmmode O \else $O$\fi}
\DeclareUnicodeCharacter{1D443}{\relax\ifmmode P \else $P$\fi}
\DeclareUnicodeCharacter{1D444}{\relax\ifmmode Q \else $Q$\fi}
\DeclareUnicodeCharacter{1D445}{\relax\ifmmode R \else $R$\fi}
\DeclareUnicodeCharacter{1D446}{\relax\ifmmode S \else $S$\fi}
\DeclareUnicodeCharacter{1D447}{\relax\ifmmode T \else $T$\fi}
\DeclareUnicodeCharacter{1D448}{\relax\ifmmode U \else $U$\fi}
\DeclareUnicodeCharacter{1D449}{\relax\ifmmode V \else $V$\fi}
\DeclareUnicodeCharacter{1D44A}{\relax\ifmmode W \else $W$\fi}
\DeclareUnicodeCharacter{1D44B}{\relax\ifmmode X \else $X$\fi}
\DeclareUnicodeCharacter{1D44C}{\relax\ifmmode Y \else $Y$\fi}
\DeclareUnicodeCharacter{1D44D}{\relax\ifmmode Z \else $Z$\fi}
\DeclareUnicodeCharacter{1D44E}{\relax\ifmmode a \else $a$\fi}
\DeclareUnicodeCharacter{1D44F}{\relax\ifmmode b \else $b$\fi}
\DeclareUnicodeCharacter{1D450}{\relax\ifmmode c \else $c$\fi}
\DeclareUnicodeCharacter{1D451}{\relax\ifmmode d \else $d$\fi}
\DeclareUnicodeCharacter{1D452}{\relax\ifmmode e \else $e$\fi}
\DeclareUnicodeCharacter{1D453}{\relax\ifmmode f \else $f$\fi}
\DeclareUnicodeCharacter{1D454}{\relax\ifmmode g \else $g$\fi}
\DeclareUnicodeCharacter{1D456}{\relax\ifmmode i \else $i$\fi}
\DeclareUnicodeCharacter{1D457}{\relax\ifmmode j \else $j$\fi}
\DeclareUnicodeCharacter{1D458}{\relax\ifmmode k \else $k$\fi}
\DeclareUnicodeCharacter{1D459}{\relax\ifmmode l \else $l$\fi}
\DeclareUnicodeCharacter{1D45A}{\relax\ifmmode m \else $m$\fi}
\DeclareUnicodeCharacter{1D45B}{\relax\ifmmode n \else $n$\fi}
\DeclareUnicodeCharacter{1D45C}{\relax\ifmmode o \else $o$\fi}
\DeclareUnicodeCharacter{1D45D}{\relax\ifmmode p \else $p$\fi}
\DeclareUnicodeCharacter{1D45E}{\relax\ifmmode q \else $q$\fi}
\DeclareUnicodeCharacter{1D45F}{\relax\ifmmode r \else $r$\fi}
\DeclareUnicodeCharacter{1D460}{\relax\ifmmode s \else $s$\fi}
\DeclareUnicodeCharacter{1D461}{\relax\ifmmode t \else $t$\fi}
\DeclareUnicodeCharacter{1D462}{\relax\ifmmode u \else $u$\fi}
\DeclareUnicodeCharacter{1D463}{\relax\ifmmode v \else $v$\fi}
\DeclareUnicodeCharacter{1D464}{\relax\ifmmode w \else $w$\fi}
\DeclareUnicodeCharacter{1D465}{\relax\ifmmode x \else $x$\fi}
\DeclareUnicodeCharacter{1D466}{\relax\ifmmode y \else $y$\fi}
\DeclareUnicodeCharacter{1D467}{\relax\ifmmode z \else $z$\fi}

\DeclareUnicodeCharacter{1E67}{\.{\v s}}
\DeclareUnicodeCharacter{1E11}{\relax\ifmmode \c{d} \else $\c{d}$\fi}
\DeclareUnicodeCharacter{1ECB}{\relax\ifmmode \d{i} \else $\d{i}$\fi}
\DeclareUnicodeCharacter{1D8D}{\relax\ifmmode \textlhookx \else $\textlhookx$\fi}
\DeclareUnicodeCharacter{104}{\relax\ifmmode \k{A} \else $\k{A}$\fi}
\DeclareUnicodeCharacter{211E}{\relax\ifmmode \textrecipe \else $\textrecipe$\fi}
\DeclareUnicodeCharacter{29D}{\relax\ifmmode \textctj \else $\textctj$\fi}

\DeclareUnicodeCharacter{1E2E}{\'{\"I}}
\DeclareUnicodeCharacter{23F}{\textrts}

\DeclareUnicodeCharacter{2C73}{\varw}

\DeclareUnicodeCharacter{2127}{\mho}

\DeclareUnicodeCharacter{28C}{\textturnv}
\DeclareUnicodeCharacter{252}{\textturnscripta}
\DeclareUnicodeCharacter{259}{\schwa}
\DeclareUnicodeCharacter{25B}{\m{e}}
\DeclareUnicodeCharacter{266}{\m{h}}
\DeclareUnicodeCharacter{127}{\B{h}}
\DeclareUnicodeCharacter{27E}{\textfishhookr}
\DeclareUnicodeCharacter{281}{\textinvscr}

\graphicspath{{./Definitions/}{./figs/}{./}}

\continuouspages{}
\firstpage{1}
\pubvolume{0}
\issuenum{0}
\elocationid{0}
\articlenumber{0}
\pubyear{2026}
\doiyear{2026}
\copyrightyear{2026}
\datereceived{}
\dateaccepted{}
\dateonlinefirst{}
\datepublished{}

\usepackage{algorithm}
\usepackage{algpseudocode}
\usepackage{amssymb}

\newcolumntype{C}{>{\centering\arraybackslash}X}
\usepackage{xcolor}
\makeatletter
\@ifundefined{linenumberfont}{}{
  
}

\renewcommand{\linenumbers}{}
\makeatother

\usepackage{fancyhdr}
\fancypagestyle{plain}{
    \fancyhf{}
    \fancyfoot[L]{\footnotesize \textit{Preprint version. Submitted to arXiv.}}
    \fancyfoot[R]{\thepage}
    
}
\Title{Cost-Aware Hierarchical Multi-Agent Ransomware Detection and Family Attribution}

\Author{Mubashar Iqbal\textsuperscript{1,2}, Hifsa Asif\textsuperscript{1}, Saddam Hussain Khan\textsuperscript{4}, Umme Zahoora\textsuperscript{5} and Asifullah Khan\textsuperscript{1,2,3,*}}

\AuthorNames{Mubashar Iqbal, Hifsa Asif, Saddam Hussain Khan, Umme Zahoora and Asifullah Khan}

\address{%
\textsuperscript{1}Pattern Recognition Lab, DCIS, Pakistan Institute of Engineering and Applied Sciences (PIEAS), Nilore, Islamabad, Pakistan

\textsuperscript{2}Deep Learning Lab, Center for Mathematical Sciences, PIEAS, Nilore, Islamabad, Pakistan

\textsuperscript{3}PIEAS Artificial Intelligence Center (PAIC), Nilore, Islamabad, Pakistan

\textsuperscript{4}Interdisciplinary Research Center for Smart Mobility and Logistics, King Fahd University of Petroleum \& Minerals, Dhahran 31261, Saudi Arabia

\textsuperscript{5}Department of Creative Technology, AIR University, E9 Campus, Islamabad, Pakistan.
}%

\corres{Corresponding Author: Asifullah Khan. Email: asif@pieas.edu.pk}

\abstract{Sandbox execution and memory forensics are among the most resources intensive stages of malware triage. Static analysis can scale to millions of files while dynamic and memory analysis require minutes per sample. Because multimodal ransomware detectors often apply every modality to every sample, costs and latency rise even when static evidence suffice.We proposea cost aware Hierarchical Multi Agent System (HMAS) that frames evidence acquisition as a budgeted sequential decision problem enabling timely ransomware detection. Specialist agents generate schema validated risk signals for each modality while domain controllers aggregate evidence and coordinate analysis. The orchestrator starts with static evidence and escalates only if confidence is low or signals disagree. A locally hosted language model provides a limited, accoutable review without replacing the deterministic pipeline. Experiments on 16 ransomware families show 94.12\% binary accuracy (F1 0.9376, ROC-AUC 0.9989). For multi class family attribution, it achieves 98.90\% accuracy (macro-F1 0.9748). The system resolves 57.95\% of the samples using static analysis alone whhile only 6.21\% require the full pipeline.
Compared with exhaustive cost of 12.0 units,the proposed system reduces te average cost to 6.724 units, representing approximately 44\% saving.  Zero day experiments further examine generalization to unseen ransomware families. Additional evaluations examine cost sensitivity and component level contribution. LLM-assisted review provides only marginal improvement in F1 while introducing additional computational overhead. Overall, the results show the potential of adaptive hierarchical multi agent orchestration to balance ransomware detection performance, family attribution analysis cost and decision latency.}

\begin{document}

\section{Introduction}
\label{sec:intro}

Ransomware is still one of the biggest threat classes of malware due to its direct impact on data availability \cite{ref-1,ref-2}. Security analysts process a large volume of sample streams. They are required to make a decision for every sample (i) whether 
it is ransomware (ii) the family to which it belongs as it will determine a response playbook \cite{ref-3}. 
For any malware sample, there are three types of evidence that are available at differing costs and delays. The cost of static analysis of sample is low and it causes less delay. 
Dynamic analysis is the process of running the sample in a sandbox and is prone to evasion \cite{ref-4}. Memory 
forensics is used to capture volatile artifacts like injected code and unpacked payloads but it also needs acquisition and 
offline processing incuring additional cost and delay\cite{ref-5,ref-6}. Recent detectors are able to enhance the accuracy by integrating these modalities \cite{ref-7,ref-8} but they treat all modalities equally and do not consider cost of acquisition of modalities or latency caused after execution of each modality. 
In operational environments dynamic and memory analysis increase time to triage a sample. therefore, question is not just which model is most accurate but also how much costly and delayed evidence is obtained in a certain model. So it becomes a sequential classification problem  \cite{ref-9,ref-10,ref-11}. It is widely studied in machine learning but not explored in ransomware detection and family attribution. 

\subsection{Problem statement}
Let a sample $x$ have evidence available through modalities $\mathcal{M}=\{S,D,M\}$ (static, dynamic, memory)
with acquisition costs $C_S \ll C_D \lesssim C_M$. A strategy $A\subseteq\mathcal{M}$ yields evidence $E_A$, a
detection decision $D$ and, for ransomware, a family $F$. We seek a per-sample strategy that maximizes decision
utility net of acquisition cost,
\begin{equation}
A^{*}(x) = \arg\max_{A}\big[\, U(D,F \mid E_A) - \lambda\, C(A) \,\big],
\label{eq:objective}
\end{equation}
where $\lambda$ sets the exchange rate between reliability and cost. Because $U$ depends on the unseen label,
$A^*$ cannot be computed directly; it must be approximated from the current evidence state (Section~\ref{sec:formulation}).

\subsection{Gap and contributions}
Prior ransomware work optimizes detection or attribution accuracy for a fixed modality set
\cite{ref-12,ref-13,ref-14,ref-15,ref-7}. Cascaded and budgeted classifiers
\cite{ref-10,ref-11,ref-16} decide when to stop, but have not been studied for ransomware with
specialis agent structure and family attribution. Multi agent frameworks for cyber security usually follow fixed workflows
\cite{ref-17,ref-18,ref-19}. We list our contributions as under.
\begin{enumerate}
\item A cost aware hierarchical multi agent architecture in which provenance preserving outputs make every decision auditable.
\item An escalation policy driven by calibrated confidence and within controller disagreement with an internal unit acquisition cost model and separately reported decision latency.
\item A bounded LLM review layer restricted to a three action set and evaluated by ablation
\item An empirical evaluation on the available train test split and leave one out and standalone held out family (zero day) testing
\end{enumerate}
Proposed HMAS is traiaged efficient and we test whether it preserves detection quality at lower acquisition cost.

\subsection{Relation to our earlier work}
Our earlier multimodal multi agent ransomware study \cite{ref-19} fuses static, dynamic and network agents
through a learned transformer to maximize family attribution accuracy, acquiring all modalities for every sample.
Proposed HMAS differs in objective, mechanism and evaluation.

The paper is organized as follows. Section~\ref{sec:related} reviews related work. In Section~\ref{sec:formulation}
We formulized problem and theat model. We discuss proposed cost aware HMAS in Section~\ref{sec:framework}. Section~\ref{sec:impl} details
the implementation. Section~\ref{sec:method} discusses the evaluation protocol. In Section~\ref{sec:results} the results are shown. 
Section~\ref{sec:discussion} includes the discussion and threats to validity and Section~\ref{sec:conclusion} concludes our work on HMAS.

\section{Related Work}
\label{sec:related}

Recent surveys highlight the use of machine learning techniques instead of traditional signature based approaches for ransomware detection . However, ransomware family attribution is still challenging due to variations among samples of same family \cite{ref-2,ref-20}. Authors have used explainable
dynamic analysis models in \cite{ref-12}, early detection from pre attack API calls \cite{ref-14},
zero-shot early detection \cite{ref-13}, transformer models in combination with assembly language functions \cite{ref-15}
and fused byte alongwith API features \cite{ref-7}. Cost sensitive ensembles for imbalanced ransomware
data is also proposed in \cite{ref-21}. Further, structured heat map learning for multi family classification is explored in \cite{ref-22}.
Iqbal et al. \cite{ref-23} combined signature based techniques with API driven behavioral analysis
for early ransomware detection, while Ghourabi and Chouaib \cite{ref-24} employed LLM embeddings
and ensemble learning to improve detection. LLM assisted pre training has also been applied to
ransomware semantics \cite{ref-25}. Although these approaches improve detection through complementary classifiers, inference time selection of analysis modalities
under explicit acquisition budgets is not explored well.

Fusing static and dynamic features improves robustness \cite{ref-8}. Steadman et al.
\cite{ref-26} showed the benefits of combining visual and structural representations
for polymorphic malware detection, while Wu et al. \cite{ref-27} proposed multi source feature
fusion using behavioral information and self attention. These studies support the value of
complementary evidence but fusion alone does not establish whether every modality
must be acquired for each sample. Fixed pipelines can therefore incur unnecessary analysis
costs on easy cases. Robustness analyses of full detection system further show that
operational choices around the classifier matter as much as the classifier itself
\cite{ref-28}.

Process level volatile memory features support ransomware detection \cite{ref-5} and LLMs
combined with Volatility can triage ransomware processes in memory dumps \cite{ref-6}.
Memory evidence is particularly valuable when other modalities are inconclusive but
its acquisition and processing costs mandates selective use to avoid cost and reduce latency in decisions.

Cascaded classifier also trade accuracy against feature evaluation cost
\cite{ref-10,ref-9} adaptive gating routes easy instances to inexpensive models
\cite{ref-11} and deep reinforcement learning can acquire costly features sequentially
\cite{ref-16}. Selective classification abstains when confidence is low \cite{ref-29},
while calibration \cite{ref-30} and conformal prediction \cite{ref-31} provide foundations
for assessing prediction confidence. HMAS draws on these principles for ransomware triage,
where the ``features'' correspond to entire analysis modalities with substantially
different acquisition costs.

LLM agents and multi-agent frameworks are increasingly applied to security tasks
\cite{ref-32}, including hierarchical planning with task-specific agents \cite{ref-33},
hierarchical multi-agent AI for cyber-physical systems \cite{ref-17}, hierarchical
multi-agent reinforcement learning for network defense \cite{ref-18}, and GPT-4-assisted
malware dynamic analysis \cite{ref-34}. Alsubait \cite{ref-35} further combined
LLM-based threat analysis with multi-agent reinforcement learning for autonomous
cybersecurity detection and response. These developments demonstrate the potential
of coordinated agents for security analysis, but agent coordination and cost-aware
evidence acquisition address distinct operational requirements. Our framework uses
AutoGen \cite{ref-36} only for the reviewer; evidence generation and routing remain
deterministic.

Malware classification results can be inflated by spatial and temporal bias
\cite{ref-37}, while concept drift degrades family classifiers \cite{ref-38}.
Common pitfalls in security machine learning \cite{ref-39} include data leakage
and unrealistic evaluation splits. We therefore identify family-disjoint and
temporally separated evaluations as important follow-up protocols alongside
the reported random-split results.

Overall, existing research has advanced ransomware detection, multimodal feature
fusion, memory forensics, and agent-based cybersecurity analysis. However, their
combination does not necessarily provide an inference-time mechanism for balancing
evidence acquisition cost, classification confidence, and family attribution.
HMAS addresses this integration problem through hierarchical routing across
static, dynamic, and memory analysis, with optional LLM-assisted review and
explicit cost-aware evaluation. Table~\ref{tbl:positioning} positions HMAS
against representative approaches.

\begin{table}[H]
\caption{Positioning of HMAS. ``Budget-aware'' means the method decides at inference time which evidence to acquire.}
\label{tbl:positioning}
\small
\begin{tabularx}{\textwidth}{@{}p{0.2\linewidth}p{0.17\linewidth}p{0.14\linewidth}p{0.11\linewidth}p{0.13\linewidth}p{0.13\linewidth}@{}}
\toprule
Method & Modalities & Budget-aware & Family attrib. & Memory evidence & Audit trace\\
\midrule
XRan \cite{ref-12} & Dynamic & No & No & No & Explanations\\
RansoGuard \cite{ref-14} & Dynamic (early) & Partially (early stop) & No & No & No\\
Pulse \cite{ref-15} & Static (assembly) & No & No & No & No\\
RansomFormer \cite{ref-7} & Static + dynamic & No & No & No & No\\
volGPT \cite{ref-6} & Memory & No & Process-level & Yes & LLM text\\
MMMA-RA \cite{ref-19} & Static + dynamic + network & No & Yes & No & Agent trace\\
\textbf{HMAS (this work)} & Static + dynamic + memory & \textbf{Yes} & \textbf{Yes} & \textbf{Yes} & \textbf{Typed trace}\\
\bottomrule
\end{tabularx}
\end{table}

\section{Problem Formulation and Threat Model}
\label{sec:formulation}

Let $Y_d\in\{0,1\}$ be the binary label (1 = ransomware) and $Y_f\in\mathcal{F}$ the family for ransomware
samples. Given acquired evidence $E_A$ the system outputs $\hat Y_d=f_d(E_A)$ and $\hat Y_f=f_f(E_A)$.

\subsection{Measured cost model}
\label{sec:cost}
For modality $i$ we define
\begin{equation}
C_i = \alpha\, T_i + \beta\, R_i + \gamma\, Q_i,
\label{eq:cost-i}
\end{equation}
where $T_i$ denotes acquisition plus analysis time in the cost model definition, $R_i$ is the
resource weighted compute footprint (CPU seconds and memory seconds, normalized), and $Q_i$ is the
queueing/occupancy time of a scarce resource (a sandbox VM slot or memory acquisition host). For a strategy $A$,
$C(A)=\sum_{i\in A}C_i$, and cost reduction relative to exhaustive analysis is
\begin{equation}
CR = \frac{C_{\text{exh}}-C_{\text{adaptive}}}{C_{\text{exh}}}\times 100.
\label{eq:cr}
\end{equation}
The reported comparative acquisition costs are internal units rather than direct measurements of wall clock acquisition time or analyst effort. Decision latency is reported separately.

\subsection{Calibrated confidence, stopping, and escalation}
\label{sec:stopping}
After each modality the controller yields a risk score $s\in[0,1]$. We calibrate scores per modality on a
validation split and define confidence $\mathrm{Conf}=\max(s,1-s)$ and
predictive entropy $H$. Acquisition stops when
\begin{equation}
\mathrm{Conf}\ge\tau_c\;\;\wedge\;\; H\le\tau_H\;\;\wedge\;\; \sigma_{\text{agents}}\le\tau_\sigma,
\label{eq:stop}
\end{equation}
where $\sigma_{\text{agents}}$ is the standard deviation of child agent risk scores inside the controller
(disagreement). Otherwise the orchestrator considers the next modality. The expected utility criterion
\begin{equation}
EU_i=\mathbb{E}[\Delta Q_i]-\lambda C_i
\label{eq:eu}
\end{equation}
motivates the design, with $\Delta Q_i$ the expected gain in decision quality from modality $i$. In the
\emph{implemented} policy, Eq.~\eqref{eq:eu} is approximated by threshold rules on $(\mathrm{Conf},H,\sigma)$
whose values, together with $\lambda$, are tuned on a validation split only (Section~\ref{sec:method}); we compare against a
learned value of information gate as a stronger alternative in the ablation.

\subsection{Threat model}
\label{sec:threat}
The adversary controls the sample. It may pack or obfuscate code (degrading static evidence), detect sandboxes or
delay execution (degrading dynamic evidence), and tamper with memory artifacts (degrading memory evidence). We
additionally consider a routing aware adversary who knows that confident static scores short circuit
analysis and perturbs static features to fall below the benign short circuit threshold, thereby suppressing
escalation. Infrastructure compromise and prompt injection against the LLM
reviewer are out of scope. The reviewer receives only structured, schema validated agent outputs and its
output space is restricted to three actions. 

\section{Cost Aware Hierarchical Multi Agent Framework}
\label{sec:framework}
\begin{figure}[t]
\centering
\includegraphics[scale=0.7]{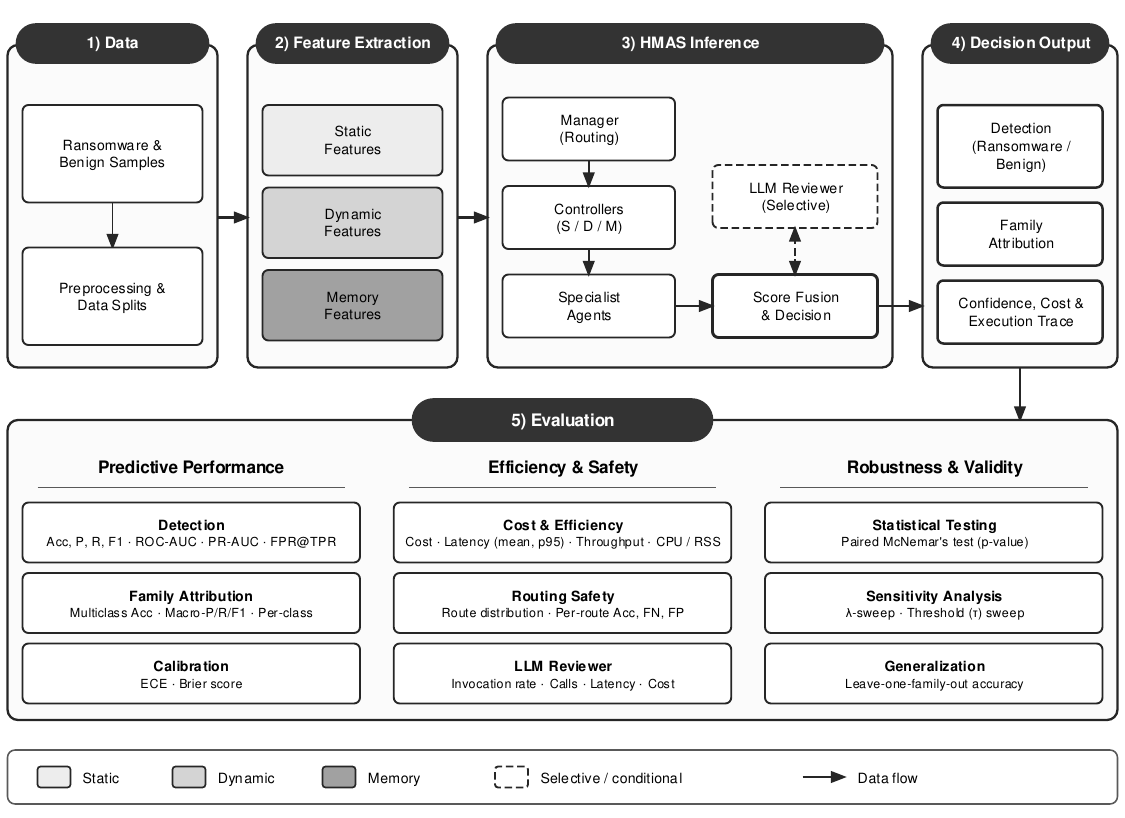}
\caption{HMAS architecture: feature extraction pipeline (left), hierarchical decision pipeline (right), and the optional LLM reasoning layer.}
\label{fig:pipeline}
\end{figure}

\begin{figure}[H]
\centering
\includegraphics[scale=0.7]{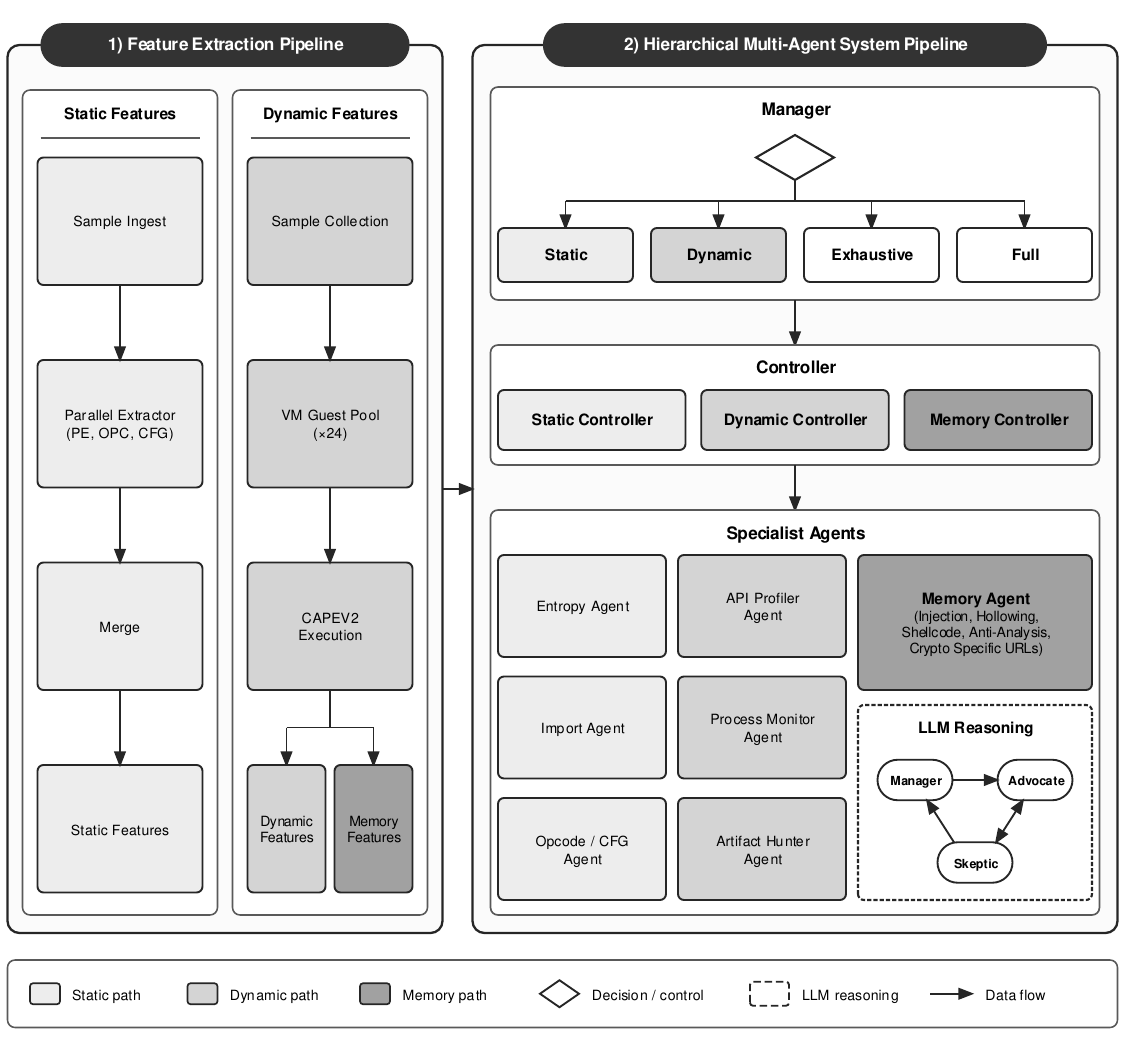}
\caption{HMAS architecture: feature extraction pipeline (left), hierarchical decision pipeline (right), and the optional LLM reasoning layer.}
\label{fig:hmas_architecture}
\end{figure}

\begin{figure}[!t]
\centering
\includegraphics[scale=0.7]{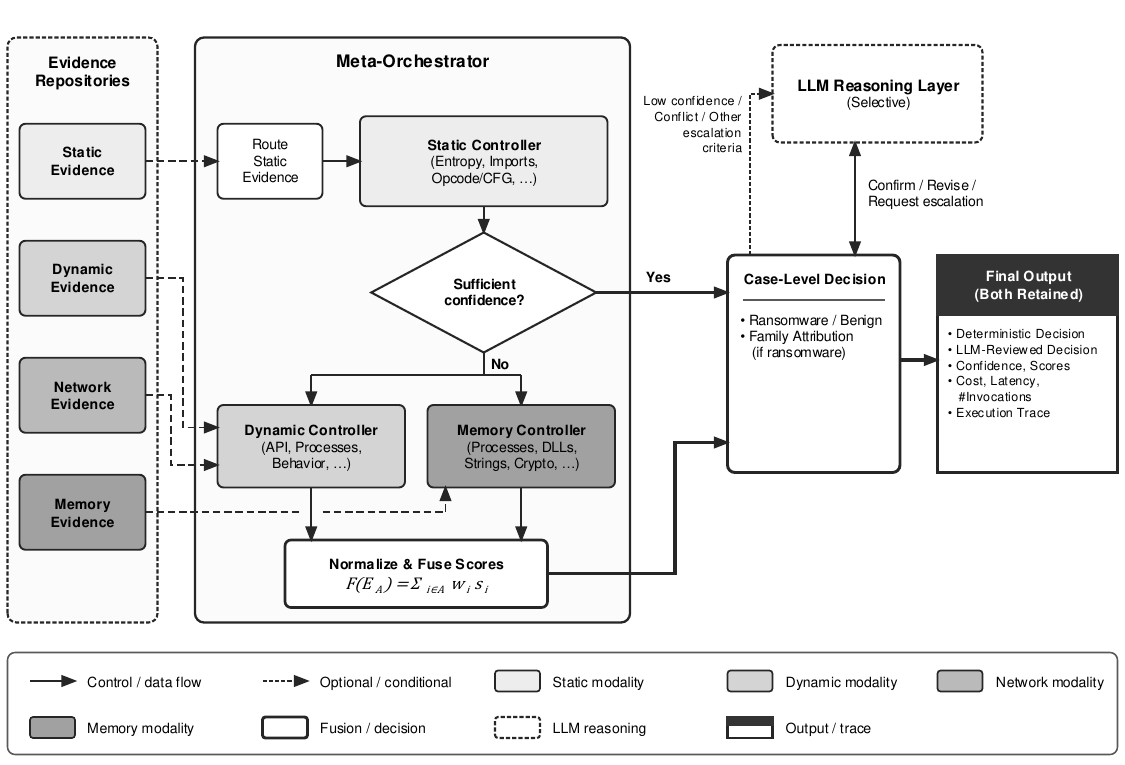}
\caption{Runtime pipeline: static evidence is processed first; the Meta-Orchestrator escalates on insufficient confidence or disagreement, fuses the acquired scores, and invokes the bounded reviewer.}
\label{fig:meta_orchastrator}
\end{figure}

Fig.~\ref{fig:pipeline} gives an end to end overview of the proposed framework, from data preparation and multimodal feature extraction through HMAS inference and decision output to the final evaluation. The following subsections describe each stage in turn.
The HMAS has four levels (Fig.~\ref{fig:hmas_architecture}). A Level-0 Meta Orchestrator; Level-1 domain
controllers (static, dynamic, memory), Level-2 specialist agents and Level-1.5 LLM reviewer. The
runtime decision flow is shown in Fig.~\ref{fig:meta_orchastrator}.

At level 2, each agent returns an immutable, schema validated record  that contains agent name, sample ID, model version, risk score in
$[0,1]$, latency and evidence fields. Static agents are based on Entropy, Import Profiler and Opcode/CFG. Dynamic agents consists of API
Profiler, Process Monitor and Artifact Hunter. Memory agent reports memory risk, cryptographic material, C2 URLs,
unpacked payload hashes and anomaly scores.

At level 1, controllers aggregate child scores with fixed weights (static: entropy 0.30, imports 0.40, opcode/CFG 0.30;
dynamic: API 0.35, process 0.40, artifact 0.25; memory: weight 1.0), execute children concurrently, and raise an
escalation flag when child scores disagree (Eq.~\ref{eq:stop}). We compare these fixed weights against learned
weights in the ablation.

The orchestrator begins with static evidence. A static score below 0.20
terminates as benign and above 0.70 as malicious. On the other hand, ambiguous scores or controller disagreement
trigger escalation to dynamic and then memory evidence if available. Acquired scores are combined by
weighted fusion renormalized over the executed modalities. Then fused score maps to verdict bands
(benign, suspicious, malicious, ransomware) and assigned a binary label.
Algorithm~\ref{alg:hmas} summarizes the procedure.

For severe conflict cases, a locally hosted small LLM runs a Manager/Advocate/Skeptic exchange over the
structured evidence at Level 1.5. It returns one of three actions, producing a reviewed binary verdict $\hat{Y}_d^{(1)}$ that is
retained alongside. Its invocation rate, override rate and net effect are also measured and reported.

\begin{algorithm}[t]
\caption{Cost aware hierarchical decision procedure}
\label{alg:hmas}
\small
\begin{algorithmic}[1]
\Require Sample $x$; available domains $\mathbf{D}$; controllers $\mathbf{C}$; family classifier $F$; optional reviewer $R$
\Ensure Binary verdict, family, confidence, audit trace
\State $E\gets\emptyset$; $Q\gets\emptyset$; $d\gets$ static if available else first available
\Repeat
  \State Run $C_d$ on $x_d$ then collect child outputs. Also compute calibrated $s_d$, $\mathrm{Conf}$, $H$, $\sigma$
  \State $E\gets E\cup\{C_d\text{ output}\}$; $Q\gets Q\cup\{d\}$
  \If{Eq.~\eqref{eq:stop} holds \textbf{or} no domain remains} \textbf{break}
  \Else\ $d\gets$ next available domain
  \EndIf
\Until{stopped}
\State $s_{\text{final}}\gets \text{weighted\_fusion}(E)$ renormalized over $Q$; map to verdict band
\State Deterministic binary verdict $\hat{Y}_d^{(0)}$; if ransomware, apply $F$ to $E$ to obtain the
family label, else family $\gets$ Benign
\If{severe conflict}
  \State Query $R$ with structured evidence; accept only \textsc{Confirm}, \textsc{EscalateOneTier}, \textsc{DowngradeOneTier}
  \State Record reviewed binary verdict $\hat{Y}_d^{(1)}$ separately \Comment{family label is \emph{not} recomputed}
\EndIf
\State \Return $\hat{Y}_d^{(0)}$, $\hat{Y}_d^{(1)}$ (if $R$ enabled), family label, scores, $Q$, measured costs and
latencies, trace
\end{algorithmic}
\end{algorithm}
The family classifier $F$ operates exclusively on the pre review evidence $E$ and the deterministic verdict $\hat{Y}_d^{(0)}$.
Typed outputs (ControllerOutput, FusionOutput and HmasCaseOutput) preserve intermediate agent outputs.
It enabls each final decision to be traced to the underlying agent evidence, routing and escalation decisions, computational cost and latency. When review is invoked, the reviewer prompt, response and resulting decisions are also retained.
\section{Implementation}
\label{sec:impl}
\begin{figure}[H]
\centering
\includegraphics[width=.9\textwidth]{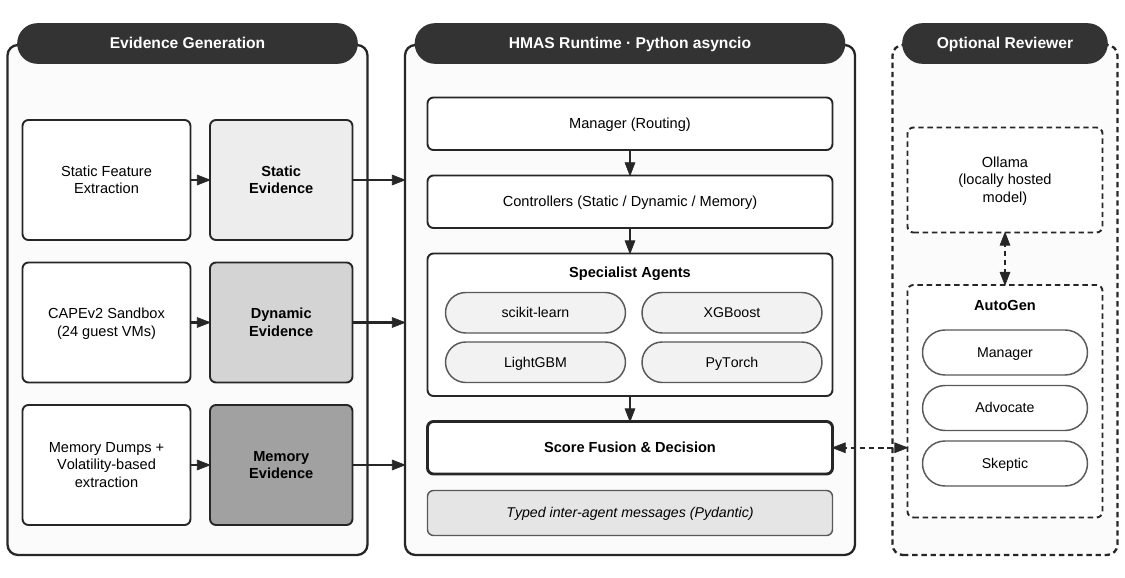}
\caption{HMAS implementation stack: evidence generation (CAPEv2 sandbox with 24 guest VMs), the asynchronous Python runtime with Pydantic-typed messages and scikit-learn, XGBoost, LightGBM and PyTorch specialist models, and the AutoGen~\cite{ref-36} reviewer on a local Ollama-hosted model (Table~\ref{tbl:llm}).}
\label{fig:sw_stack}
\end{figure}

HMAS is implemented in Python with an asynchronous execution model. Typed messages use Pydantic,
specialist models use scikit learn, XGBoost, LightGBM and PyTorch. The LLM reviewer uses AutoGen
\cite{ref-36} with a locally hosted model via Ollama (Table~\ref{tbl:llm}). Dynamic evidence is generated in a CAPEv2
sandbox with a pool of 24 guest VMs, and memory evidence from memory dumps processed with Volatility based
extraction (Fig.~\ref{fig:sw_stack}).
\begin{table}[H]
\caption{Reviewer LLM configuration.}
\label{tbl:llm}
\small
\begin{tabularx}{\textwidth}{@{}p{0.3\linewidth}p{0.5\linewidth}@{}}
\toprule
Item & Setting\\
\midrule
Backend & Ollama + AutoGen (\texttt{AssistantAgent}, \texttt{RoundRobinGroupChat})\\
Models & Phi-3 Mini; Qwen2.5-3B\\
Context / output & 4096 / 512 tokens\\
Sampling & $T{=}0.1$, top-$p{=}0.85$, top-$k{=}20$, repetition penalty 1.05, seed 42\\
Input / output & Structured agent evidence / one of three actions\\
\bottomrule
\end{tabularx}
\end{table}

\section{Experimental Protocol}
\label{sec:method}

\subsection{Dataset and splits}
\begin{figure}[H]
\centering
\includegraphics[width=.7\textwidth]{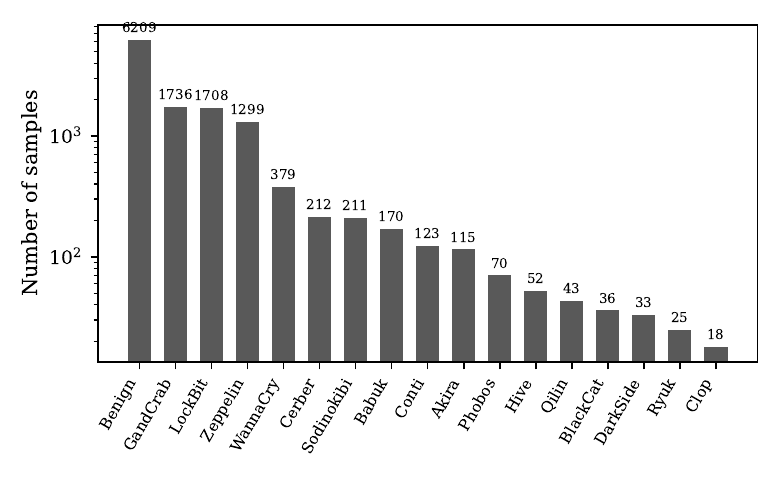}
\caption{Class distribution (16 ransomware families and benign).}
\label{fig:families}
\end{figure}

Experiments were performed using a subset of the RanDS dataset \cite{ref-40}, an open collection containing 104,616 ransomware samples representing 533 families and 110,788 benign samples targeting Windows Portable Executable (PE) files. From this collection, 8{,}013 ransomware samples
from 16 families and 8{,}015 benign samples (Fig.~\ref{fig:families}) were selected for the study. Static, dynamic and memory features are extracted as described in \cite{ref-42}. Family sizes are strongly imbalanced. We evaluate the system under two protocols.
(P1) Random: stratified 70/10/20 train/validation/test. (P2) The closest available evidence for
cross family generalization is a leave-one-family-out standalone evaluation.

\subsection{Compared policies}
\begin{itemize}
\item \textbf{Single modality}: B1 static-only, B2 dynamic-only, B3 memory-only (each with its best tuned model).
\item \textbf{HMAS}: deterministic (H0) and with LLM review (H1), SD HMAS(Fixed), SDM HMAS (Fixed)
\end{itemize}

\subsection{Metrics}
To measure performance in terms of detection  accuracy, precision, recall, F1, ROC-AUC, PR-AUC, FPR, TPR, and false negative rate are used.
Similarly for family attribution accuracy, macro precision/recall/F1, and per-family F1 are used.
For calibration of agents,  ECE \cite{ref-30} and Brier score per modality are used. To measure efficiency, we used modeled internal acquisition cost (Eq.~\ref{eq:cost-i}), cost reduction (Eq.~\ref{eq:cr}), median and 95th-percentile
time to decision, average number of acquired modalities, and \emph{throughput} (samples/hour) from a discrete event
simulation of a 24-slot sandbox pool and one memory analysis host under Poisson arrivals at several loads.

\section{Results}
\label{sec:results}

\subsection{RQ1: Detection accuracy versus acquisition cost}
\label{sec:rq1}
Table~\ref{tbl:detection} reports binary detection for single modality baseline alongwith fixed HMAS scenarios.
\begin{table*}[t]
\caption{Binary detection at full scale policies}
\label{tbl:detection}
\small
\begin{tabularx}{\textwidth}{@{}lCCCCCC@{}}
\toprule
Policy & Acc. & Prec. & Rec. & F1 & ROC-AUC & Avg.\ modalities\\
\midrule
B1 Static-only & {0.9130} & {0.9988} & {0.8273} & {0.9050} & {0.9983} & 1.00\\
B2 Dynamic-only & {0.9329} & {0.9642} & {0.8994} & {0.9307} & {0.9883} & 1.00\\
B3 Memory-only & {0.5002} & {1.0000} & {0.0005} & {0.0010} & {0.7644} & 1.00\\
\midrule
SD HMAS (Fixed) & {0.9647} & {0.9993} & {0.9302} & {0.9635} & {0.9995} & 2.00\\
SDM HMAS (fixed) & {0.9401} & {0.9993} & {0.8811} & {0.9364} & {0.9995} & 3.00\\

H0 HMAS (deterministic) & {0.9412} & {0.9993} & {0.8831} & {0.9376} & {0.9989} & {0.618}\\
H1 HMAS + LLM review & {0.9420} & {0.9474} & {0.9361} & {0.9417} & {0.9989} & {0.618}\\
\bottomrule
\end{tabularx}
\end{table*}
Two findings stand out among the fixed and single modality baselines. First, B3 (memory only) is not an informative
classifier. It predicts almost every case negative (4 true positives out of 8{,}016
ransomware cases, ROC-AUC 0.76 indicates the underlying score does rank samples somewhat. Second, SD HMAS (fixed static+dynamic fusion) has a higher F1 (0.9635) than both
SDM HMAS (exhaustive, 0.9364, which adds the uninformative memory score) and H0 (0.9376, which adds memory only when
escalation triggers). Adding memory evidence under the current fixed weight fusion rule does not help and mildly
hurts binary F1 relative to stopping at dynamic evidence  consistent with B3's near chance standalone performance.

Comparing H0 and H1 shows the reviewer's effect. Results show that enabling locally hosted LLM moves 330 previously missed ransomware cases to
correctly detected (recall $0.8831\rightarrow0.9361$) at the cost of 320 previously correct benign cases becoming
false alarms (precision $0.9993\rightarrow0.9474$). These two groups are disjoint by construction -- true-label
counts are identical between H0 and H1 ($TN{+}FP=6{,}209$, $FN{+}TP=6{,}230$ in both) -- so the reviewer is not
simply noisier than the deterministic pipeline, it is systematically biased toward escalating rather than
downgrading final decisions. The net F1 change is small (+0.0041) because the two effects largely offset. Whether this
trade is worthwhile depends on the relative operational cost of a missed ransomware sample versus a false alarm,
which this evaluation does not quantify. It comes at a steep resource cost.

In addition to detection performance, HMAS is evaluated in terms of
analysis cost to assess the efficiency of adaptive evidence acquisition.
Rather than measuring monetary expenses or execution time directly,
we employ a normalized computational cost model that assigns
predefined weights to specialist agent execution and LLM interactions.
For each analyzed sample, the total cost is defined as

\begin{equation}
C_i =
\sum_{a \in \mathcal{A}_i} w_a
+ 5N_i^{\mathrm{mgr}}
+ 5N_i^{\mathrm{rev}},
\label{eq:analysis_cost}
\end{equation}

\noindent
where $\mathcal{A}_i$ denotes the specialist agents invoked for
sample $i$, $w_a$ represents the assigned cost of agent $a$,
and $N_i^{\mathrm{mgr}}$ and $N_i^{\mathrm{rev}}$ denote the
numbers of manager and advocate/skeptic LLM turns, respectively.
Each LLM turn contributes five normalized cost units.

For comparison, static only, static plus dynamic, and exhaustive
analysis are assigned fixed costs of 4, 10, and 12 units per
sample, respectively. The average cost of a policy is computed
over the $N=12{,}439$ evaluated samples as

\begin{equation}
\overline{C} =
\frac{1}{N}\sum_{i=1}^{N} C_i.
\label{eq:average_cost}
\end{equation}

The relative cost reduction compared with exhaustive analysis is

\begin{equation}
S =
\left(
1-\frac{\overline{C}_{\mathrm{policy}}}
{\overline{C}_{\mathrm{exhaustive}}}
\right)\times 100\%.
\label{eq:cost_savings}
\end{equation}

Table~\ref{tab:cost_comparison} summarizes the average and total
modeled costs of the evaluated policies.

\begin{table}[htbp]
\centering
\caption{Comparison of normalized analysis costs across policies.}
\label{tab:cost_comparison}
\small
\setlength{\tabcolsep}{5pt}
\renewcommand{\arraystretch}{1.15}
\begin{tabular}{lccc}
\hline
\textbf{Policy} &
\textbf{Avg. cost} &
\textbf{Total cost} &
\textbf{Savings (\%)} \\
& \textbf{(units)} &
\textbf{(units)} & \\
\hline
Static-only          & 4.000 & 49,756  & 66.7 \\
Static + Dynamic     & 10.000 & 124,390 & 16.7 \\
Exhaustive Analysis  & 12.000 & 149,268 & 0.0 \\
\textbf{Full HMAS}   & \textbf{6.724} &
\textbf{83,636} & \textbf{44.0} \\
\hline
\end{tabular}
\end{table}

The adaptive routing mechanism enables HMAS to avoid unnecessary
analysis stages for samples that can be classified with sufficient
confidence using less expensive evidence. Specifically, 56.05\%
of samples are processed using static analysis alone, 39.62\%
require static and dynamic analysis, and only 4.33\% proceed
through the full analysis pipeline.

Consequently, HMAS achieves an average modeled cost of 6.724
units per sample, corresponding to approximately 44.0\% lower
cost than exhaustive analysis. Although static only analysis
has the lowest cost, it achieves lower classification performance,
as shown in Tables~\ref{tbl:detection} and
\ref{tbl:family}. HMAS combines
selective evidence acquisition with improved binary detection
and ransomware family identification, demonstrating a favorable
trade off between classification performance and modeled
analysis cost.

These results quantify savings under the predefined cost model instead of measured reductions in runtime,
energy consumption or monetary expenditure.

\begin{figure}[H]
\centering
\includegraphics[width=.8\textwidth]{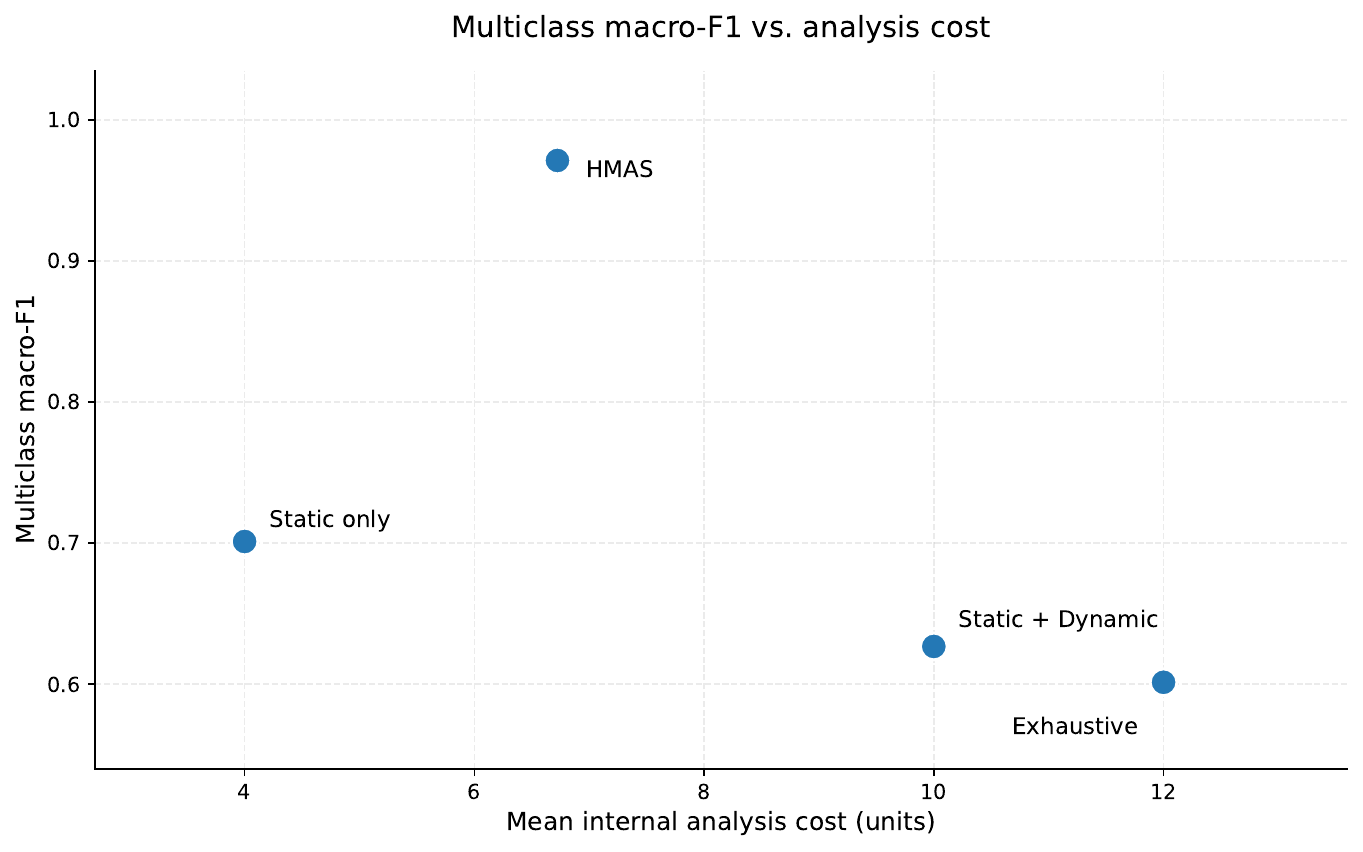}
\caption{Computational cost versus multiclass macro-F1 for HMAS and baseline analysis strategies}
\label{fig:pareto}
\end{figure}

\subsection{RQ2: Family attribution}
\label{sec:rq2}
Table~\ref{tbl:family} shows the separately evaluated 16 ransomware families and one benign. The binary and multiclass accuracy figures does not represent a single hierarchy of predictions on identical samples. Multiclass accuracy (0.9890) exceeds binary accuracy (0.9412). They are therefore interpreted as separately evaluated outputs, not as the accuracy of a single end-to-end binary-gated family verdict. A matched-prediction reconciliation is required before claiming end-to-end hierarchical accuracy.
\begin{table}[H]
\caption{Family attribution}
\label{tbl:family}
\small
\begin{tabularx}{\textwidth}{@{}lCCCC@{}}
\toprule
Policy & MC Acc. & Macro-P & Macro-R & Macro-F1\\
\midrule
B1 Static-only & {0.9076} & {0.9765} & {0.6264} & {0.7022}\\
B2 Dynamic-only & {0.8950} & {0.8893} & {0.7502} & {0.7888}\\
B3 Memory-only & {0.5001} & {0.0833} & {0.0562} & {0.0383}\\
\midrule
SDM HMAS(Fixed) & {0.9338} & {0.9806} & {0.7341} & {0.8204}\\
H0 HMAS & {0.9890} & {0.9887} & {0.9890} & {0.9748}\\
H1 HMAS + LLM review & {0.9419} & {0.9473} & {0.9361}& {0.9417}\\
\bottomrule
\end{tabularx}
\end{table}

Family attribution is where cost aware HMAS shows its clearest best supported gain over every single or fixed modality
HMAS. Macro-F1 rises from 0.70 (static only) and 0.82 (exhaustive) to 0.98 in case of deterministic low cost HMAS and every one of the 17 classes
individually exceeds 0.93 F1. The memory only
model (B3) is again uninformative for attribution (macro-F1 0.038, i.e., it essentially never assigns the correct
family) which matters because memory evidence is exactly what HMAS adds on the hardest 6.21\% of cases.
\begin{table}[h]
\centering
\caption{Comparison of per-class F1-scores across different analysis policies.}
\label{tab:multiclass_comparison}
\small
\setlength{\tabcolsep}{4pt}
\renewcommand{\arraystretch}{1.15}
\begin{tabular}{lccccc}
\hline
\textbf{Class} & \textbf{Support} &
\textbf{Static} & \textbf{S+D} &
\textbf{Exhaustive} & \textbf{HMAS} \\
\hline
Akira       & 125   & 0.252 & 0.619 & 0.580 & \textbf{0.988} \\
Babuk       & 191   & 0.155 & 0.525 & 0.508 & \textbf{0.953} \\
Benign      & 5,810 & 0.909 & 0.780 & 0.766 & \textbf{0.993} \\
BlackCat    & 37    & 0.879 & 0.679 & 0.604 & \textbf{0.972} \\
Cerber      & 223   & 0.709 & 0.578 & 0.541 & \textbf{0.956} \\
Clop        & 17    & 0.583 & 0.692 & 0.692 & \textbf{0.938} \\
Conti       & 126   & 0.573 & 0.578 & 0.554 & \textbf{0.942} \\
DarkSide    & 33    & 0.553 & 0.533 & 0.533 & \textbf{0.954} \\
GandCrab    & 1,902 & 0.975 & 0.730 & 0.664 & \textbf{0.993} \\
Hive        & 57    & 0.879 & 0.578 & 0.578 & \textbf{0.958} \\
LockBit     & 1,765 & 0.948 & 0.651 & 0.635 & \textbf{0.984} \\
Phobos      & 79    & 0.881 & 0.689 & 0.655 & \textbf{0.975} \\
Qilin       & 41    & \textbf{1.000} & 0.453 & 0.423 & \textbf{1.000} \\
Ryuk        & 26    & 0.556 & 0.632 & 0.632 & \textbf{0.939} \\
Sodinokibi  & 219   & 0.947 & 0.661 & 0.627 & \textbf{0.984} \\
WannaCry    & 396   & 0.150 & 0.637 & 0.613 & \textbf{0.997} \\
Zeppelin    & 1,392 & 0.972 & 0.641 & 0.618 & \textbf{0.983} \\
\hline
\end{tabular}

\vspace{2pt}
\footnotesize{
S+D: Static + Dynamic.
Bold values indicate the highest F1-score for each class.
}
\end{table}
\subsubsection{Comparative Multiclass Classification Performance}

To investigate the contribution of HMAS, it was compared
with three alternative analysis techniques.
Table~\ref{tab:hmas_comparison} summarizes
their multiclass classification performance
and computational requirements.

\begin{table}[h]
\centering
\caption{Multiclass performance
and computational efficiency of HMAS and
alternative analysis strategies}
\label{tab:hmas_comparison}
\small
\begin{tabular}{lrrrr}
\hline
\textbf{Method} &
\textbf{Acc.} &
\textbf{Macro-F1} &
\textbf{Cost} &
\textbf{Latency} \\
 & \textbf{(\%)} & & \textbf{(units)}
 & \textbf{(ms)} \\
\hline
Static only
 & 90.20 & 0.7011 & 4.00 & 15.39 \\
Static + Dynamic
 & 73.29 & 0.6267 & 10.00 & 239.29 \\
Exhaustive
 & 71.22 & 0.6013 & 12.00 & 242.78 \\
\textbf{Full HMAS}
 & \textbf{98.85} & \textbf{0.9712}
 & \textbf{6.72} & \textbf{107.11} \\
\hline
\end{tabular}
\end{table}

The proposed HMAS achieved a multiclass
macro-F1 of 0.97, compared with 0.70,
0.62 and 0.60 for static-only,
static and dynamic, and exhaustive analysis respectively.
These results correspond to absolute macro-F1
improvements of approximately 27.00, 34.45,
and 36.98 \%, respectively.
The improvement over exhaustive analysis
is calculated as
\begin{equation}
\begin{aligned}
\Delta F1 &= F1_{\mathrm{HMAS}} - F1_{\mathrm{Exhaustive}}\\
          &= 0.9712 - 0.6013 = 0.3699.
\end{aligned}
\end{equation}
The observed improvement suggests that
adaptive multi-agent orchestration provides
a more effective classification strategy
than indiscriminately executing all available
analysis stages.
In particular, HMAS combines selective
analysis execution with coordinated
decision-making, enabling the system to
avoid unnecessary processing while
maintaining strong multiclass discrimination.
However, these results are obtained from
the reported multiclass ablation evaluation,
rather than from matched zero-day tests.
Thus, they establish a comparative advantage
in the evaluated multiclass setting but
do not quantify a zero-day-specific F1
improvement over the competing methods.

\subsection{RQ3: Cost and latency}
\label{sec:rq3}
Table~\ref{tbl:cost-new} reports the internal cost unit and measured mean inference latency logged by the pipeline.

\begin{table}[H]
\caption{Internal cost units and mean/p95 inference latency, default threshold, seed 42}
\label{tbl:cost-new}
\small
\begin{tabularx}{\textwidth}{@{}lCCCC@{}}
\toprule
Policy & Mean cost (units) & Cost redn.\ vs.\ exhaustive & Mean latency (ms) & p95 latency (ms)\\
\midrule
B1 Static-only & {4.00} & {66.67\%} & {15.67} &{16.08}\\
SD HMAS (Fixed) & {10.00} & {16.67\%} & {245.52} & {255.50}\\
SDM HMAS & {12.00} & {0\%} & {249.28} & {259.29}\\
H0 HMAS & {6.65} & {44.6\%} & {111.34} & {218.30}\\
\bottomrule
\end{tabularx}
\end{table}

\begin{table}[H]
\caption{Throughput observations by agent-pool-size bucket}
\label{tbl:throughput}
\small
\begin{tabularx}{\textwidth}{@{}CCCC@{}}
\toprule
Agent pool size & $n$ (cases) & Mean latency (ms) & Throughput (samples/s)\\
\midrule
3 & {7209} & {37.47} & {26.69}\\
6 & {4457} & {213.47} & {4.68}\\
7 & {773} & {211.34} & {4.73}\\
\bottomrule
\end{tabularx}
\end{table}

The cost unit results are internally consistent with the routing distribution: $0.5795 \times 4 + 0.3583 \times 10 +
0.0621 \times 12 = 6.65$, matching the logged mean cost exactly. The throughput buckets in Table~\ref{tbl:throughput}
correspond to the same three routes (agent pool size 3 = static-only route, 6 = static+dynamic, 7 = full pipeline).
\begin{figure}[H]
\centering
\includegraphics[width=.8\textwidth]{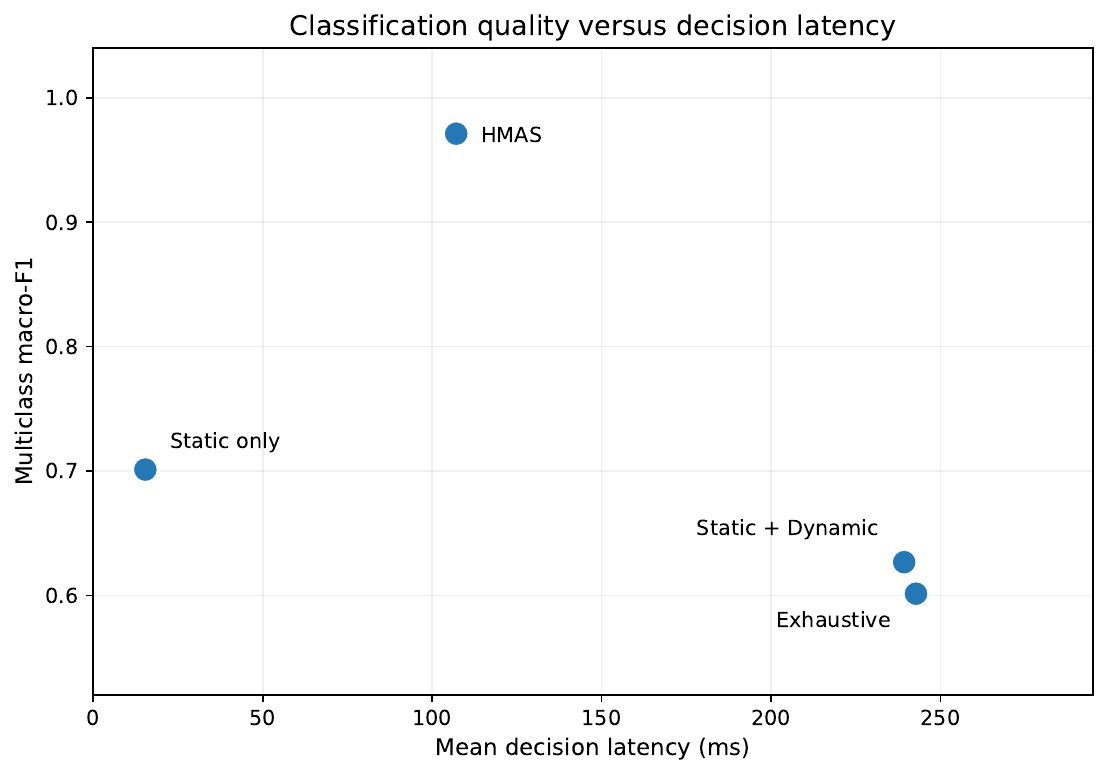}
\caption{Comparison of multiclass macro-F1 and decision latency across analysis strategies.}
\label{fig:pareto}
\end{figure}

\subsubsection{Computational Cost Reduction}

In addition to classification performance,
computational efficiency is particularly
important for intrusion detection systems
operating under limited processing resources
or strict response-time requirements.
The computational cost reduction relative
to exhaustive analysis is defined as
\begin{equation}
R_C =
\frac{C_{\mathrm{Exhaustive}} -
C_{\mathrm{HMAS}}}
{C_{\mathrm{Exhaustive}}}
\times 100,
\end{equation}
where $C$ denotes the average computational
cost in the relative units defined by
the experimental framework.
Using the measured values,
\begin{equation}
R_C =
\frac{12.00 - 6.7237}{12.00}
\times 100
= 43.97\%.
\end{equation}
Thus, HMAS reduces the average computational
cost by approximately 43.97\% relative to
exhaustive analysis.
This reduction is achieved through adaptive
selection of analysis stages. Specifically,
56.05\% of the evaluated cases were processed
using static analysis alone, 39.62\% required
combined static and dynamic analysis, and
only 4.33\% required the complete pipeline.
These routing proportions indicate that
HMAS avoids exhaustive processing for
approximately 95.67\% of the evaluated
cases.

The results suggest that selective
multi-agent orchestration can reduce
redundant computation, particularly when
the available evidence is sufficient
for an earlier decision.

\subsubsection{Decision Latency Analysis}

The reduction in computational cost was
accompanied by a decrease in average
decision latency.
HMAS required an average of 107.11 ms
per decision, compared with 242.78 ms
for exhaustive analysis.
The relative latency reduction is given by
\begin{equation}
R_L =
\frac{L_{\mathrm{Exhaustive}} -
L_{\mathrm{HMAS}}}
{L_{\mathrm{Exhaustive}}}
\times 100,
\end{equation}
where $L$  is average decision latency.
\begin{equation}
R_L =
\frac{242.78 - 107.11}{242.78}
\times 100
\approx 55.88\%.
\end{equation}
So we cay say  that HMAS achieves approximately
55.88\% lower average decision latency
as compared to exhaustive analysis.
Although static only analysis remains
the fastest configuration, its multiclass
macro-F1 is substantially lower than that
of HMAS.
This shows the trade off between
computational efficiency and classification
quality. HMAS balances multiclass
classification performance and time to decision.

The comparative ablation experiments
also show that HMAS achieves substantially
higher multiclass macro-F1 while reducing both computational cost
and decision latency relative to exhaustive analysis.
These characteristics are relevant to
machine learning based intrusion detection where detection quality must be balanced
against processing overhead and response time constraints.
Overall, the results support the
effectiveness of HMAS as a computationally
efficient multiclass ransomware analysis
framework and provide an initial assessment
of its zero day detection behavior.
\subsection{RQ4: Adaptive Routing Efficiency and Error Analysis}
\label{subsec:routing_analysis}

To evaluate the effectiveness of the proposed adaptive
orchestration mechanism, the routing decisions and
corresponding classification outcomes were analyzed
using 12,439 case level predictions.
Table~\ref{tbl:routing} summarizes the distribution
of samples across the three analysis routes, together
with their corresponding classification accuracy,
false negatives (FN), and false positives (FP).
\begin{table}[htbp]
\centering
\caption{Adaptive routing distribution and
classification outcomes for HMAS.}
\label{tbl:routing}
\small
\begin{tabular}{lrrrrr}
\hline
\textbf{Analysis route} &
\textbf{Samples} &
\textbf{Share (\%)} &
\textbf{Accuracy} &
\textbf{FN} &
\textbf{FP} \\
\hline
Static only
& 7209 & 57.95 & 0.9900 & 71 & 1 \\

Static + Dynamic
& 4457 & 35.83 & 0.9493 & 225 & 1 \\

Full pipeline
& 773 & 6.21 & 0.4386 & 432 & 2 \\
\hline
\end{tabular}
\end{table}
The results demonstrate that HMAS can substantially
reduce the requirement for exhaustive multimodal
analysis through adaptive execution.
Specifically, 57.95\% of samples were classified
using static analysis alone, while an additional
35.83\% required combined static and dynamic
analysis.
93.79\% of the evaluated samples
were processed without invoking the complete
analysis pipeline. Only 6.21\% required
additional memory based analysis.
This shows the potential
of hierarchical multi agent orchestration to reduce
computational overhead by selectively allocating
analysis resources.

The static only route achieved an accuracy of
99.00\% with 71 false negatives and one false
positive. Similarly, the static plus dynamic
achieved 94.93\% accuracy with 225 false
negatives and one false positive.
These results indicate that the earlier analysis
stages provide reliable classification outcomes.

Although the escalation mechanism concentrates
computational resources on samples requiring
additional analysis, the classification outcomes
indicate that increasing the number of analysis
modalities does not necessarily improve
decision reliability.
The weak standalone performance of memory based
analysis suggests that its contribution to
multimodal fusion requires further investigation.
In particular, confidence aware modality
weighting may help prevent weak or
uninformative features from adversely affecting
the final classification decision.

A matched evaluation of alternative fusion
strategies on the same escalated samples
would be required to isolate these effects.
Overall, the routing results demonstrate
the computational benefits of HMAS while
identifying opportunities to improve the
reliability of its final escalation stage.
These findings motivate further refinement
of modality specific confidence estimation
and adaptive fusion for robust intrusion
detection.
\subsection{RQ5: Ablation and sensitivity}
\label{sec:rq5}

Table~\ref{tbl:ablation} show an ablation study. It shows accuracy, true negatives, false positive for all samples w.r.t static only, dynamic only, memory only, adaptive routing, full HMAS with out LLM and with LLM.

\begin{table}[H]
\caption{Leave-one-component-out}
\label{tbl:ablation}
\small
\begin{tabularx}{\textwidth}{@{}lCCCCC@{}}
\toprule
Row label (as logged) & $n$ & Acc. & TN & FP & FN/TP\\
\midrule
``static'' (= B1 static-only) & 12439 & {0.9130} & 6203 & 6 & 1076/5154\\
``dynamic'' (= B2 dynamic-only) & 12439 & {0.9329} & 6001 & 208 & 627/5603\\
``memory'' (= B3 memory-only) & 12439 & {0.4994} & 6209 & 0 & 6227/3\\
$-$ adaptive routing (= SDM HMAS (Fixed)) & 12439 & {0.9401} & 6205 & 4 & 741/5489\\
$-$ LLM review (= H0) & 12439 & {0.9412} & 6205 & 4 & 728/5502\\
\bottomrule
\end{tabularx}
\end{table}

Table~\ref{tbl:lambda} shows the objective $U - \lambda C$ (Eq.~\ref{eq:objective}, accuracy minus $\lambda$ times
internal cost units). For H0 versus B1, the equality $0.9412-6.65\lambda=0.9130-4\lambda$ occurs at $\lambda\approx0.01064$. H0 has higher accuracy-adjusted utility for $0\leq\lambda<0.01064$, while B1 is preferable for $\lambda>0.01064$ under this particular objective. Thus, the selected grid values ($\lambda=0,0.1,0.5,1$) do not resolve the crossover interval. The objective measures binary accuracy and internal acquisition cost; it does not incorporate family attribution or asymmetric false-negative harm.

\begin{table}[H]
\caption{Objective value $\mathrm{Acc} - \lambda \cdot \mathrm{cost}$ at selected $\lambda$ (seed 42, default threshold).}
\label{tbl:lambda}
\small
\begin{tabularx}{\textwidth}{@{}lCCCC@{}}
\toprule
$\lambda$ & B1 Static-only & SD HMAS & SDM HMAS & H0 HMAS\\
\midrule
0.0 & {0.9130} & {0.9647} & {0.9401} & {0.9412}\\
0.1 & {0.5130} & {-0.0353} & {-0.2599} & {0.2765}\\
0.5 & {-1.0870} & {-4.0353} & {-5.0599} & {-2.3823}\\
1.0 & {-3.0870} & {-9.0353} & {-11.0599} & {-5.7058}\\
\bottomrule
\end{tabularx}
\end{table}

\subsection{RQ6: Value of the LLM reviewer}
\label{sec:rq6}
A second run enabled the LLM reviewer for the main evaluation; the separate zero-day comparison reported below did not invoke the LLM. The reviewer was invoked on 31.1\% of the 12{,}439 cases, averaging 1.48 LLM calls per invoked case. Table~\ref{tbl:llmcost}
reports its cost. The binary accuracy/precision/recall trade-off is discussed in Section~\ref{sec:rq1}
(Table~\ref{tbl:detection})
\begin{table}[H]
\caption{Cost of enabling the LLM reviewer, default threshold, single run (seed 42, $n=12{,}439$).}
\label{tbl:llmcost}
\small
\begin{tabularx}{\textwidth}{@{}lCC@{}}
\toprule
Quantity & H0 (deterministic) & H1 (LLM-enabled)\\
\midrule
Mean cost (internal units) & {6.65} & {14.07}\\
Mean decision latency & {111\,ms} & {6{,}639\,ms}\\
p95 decision latency & {218\,ms} & {22{,}751\,ms}\\
Invocation rate & -- & {31.1\%}\\
Mean LLM calls per invoked case & -- & {1.48}\\
\bottomrule
\end{tabularx}
\end{table}

Enabling the reviewer roughly doubles mean analysis cost (6.65$\to$14.07 units) and raises p95 latency by two orders
of magnitude (218\,ms$\to$22.75\,s), for a net binary F1 change of +0.0041 (Section~\ref{sec:rq1}). 

\subsection{Zero-Day Ransomware Detection and Computational Efficiency}
\label{subsec:zeroday_hmas}

\subsubsection{Zero-Day Evaluation Protocol}

To investigate the effectiveness of the proposed
Hierarchical Multi-Agent System against
previously unseen ransomware, experiments were
conducted across 16 ransomware families.
Each family was considered separately in the
zero-day evaluation, allowing the detection
capability to be examined under different
ransomware characteristics and sample distributions.
The standalone held-out-family evaluation focuses on two complementary
aspects: (i) the ability to detect ransomware
samples belonging to previously unseen families,
and (ii) the classification performance and
computational efficiency of HMAS relative to
alternative analysis strategies in a separate comparative experiment. 

\subsubsection{Detection of Previously Unseen Ransomware}

Table~\ref{tab:zeroday_results} shows the
binary detection performance obtained for
16 zero day ransomware families.
The results reveal variation
in detection capability across families. 
\begin{table}[htbp]
\centering
\caption{Standalone zero day binary detection performance
across previously unseen ransomware families.}
\label{tab:zeroday_results}
\small
\begin{tabular}{lrrr}
\hline
\textbf{Family} &
\textbf{Samples} &
\textbf{Recall (\%)} &
\textbf{F1} \\
\hline
Akira       & 399   & 33.58 & 0.503 \\
Babuk       & 600   & 30.17 & 0.464 \\
BlackCat    & 128   & 39.84 & 0.570 \\
Cerber      & 730   & 36.30 & 0.533 \\
Clop        & 58    & 29.31 & 0.453 \\
Conti       & 449   & 35.41 & 0.523 \\
DarkSide    & 111   & 28.83 & 0.448 \\
GandCrab    & 6204  & 58.64 & 0.739 \\
Hive        & 188   & 43.09 & 0.602 \\
LockBit     & 6110  & 60.67 & 0.755 \\
Phobos      & 262   & 54.58 & 0.706 \\
Qilin       & 149   & 63.76 & 0.779 \\
Ryuk        & 85    & 42.35 & 0.595 \\
Sodinokibi  & 779   & 52.76 & 0.691 \\
WannaCry    & 1311  & 27.61 & 0.433 \\
Zeppelin    & 4695  & 62.75 & 0.771 \\
\hline
\end{tabular}
\end{table}

Among the evaluated families Qilin achieved
the highest binary F1-score of 0.779, followed
by Zeppelin (0.771) and LockBit (0.755).
GandCrab and Phobos also demonstrated relatively
strong detection performance achieving F1-scores
of 0.73 and 0.70 respectively.
On the other hand WannaCry, DarkSide and Clop
exhibited comparatively lower detection
performance with F1-scores of 0.43, 0.44
and 0.45 respectively.
The unweighted macro-average binary recall across the
16 families was approximately 43.73\%,
whereas the sample-weighted recall was
approximately 55.07\%. This difference
reflects the unequal distribution of samples
among the evaluated ransomware families.
These findings demonstrate that the detection
capability varies substantially across unseen
families. Although several families exhibit
good results, the lower recall
observed for other families indicates that
zero day generalization remain challenging.
Furthermore the zero day reports assign the
generic label zeroday detected to the
held out samples. Therefore, the reported
perfect unknown class classification scores
cannot be interpreted as evidence of successful
identification of the exact unseen ransomware
family. Accordingly, binary detection recall
and F1 are used as the primary zero-day
performance indicators.

\section{Discussion and Threats to Validity}
\label{sec:discussion}

The experimental findings show the potential of the
proposed hierarchical multi agent system (HMAS) to balance
ransomware detection performance, family attribution and
computational efficiency through adaptive evidence acquisition.
HMAS selectively escalates analysis based on the available evidence and
routing decisions thus minimizing time to final decision in a critical system in case of a possible intrusion.

HMAS achieves higher binary F1 than exhaustive analysis with low cost and less decision latency. Its main advantage is shown in multiclass ransomware family attribution.
HMAS achieves a macro-F1 of 0.9748, compared with 0.7022--0.8204 for the evaluated fixed configurations 
(Table~\ref{tbl:family}). This corresponds to an absolute improvement of approximately 15.44--27.26 \% highlighting the effectiveness of adaptive
multi agent coordination for family level ransomware classification.

The routing analysis also shows the benefits of selective modality execution in terms of cost and latency.
HMAS resolves approximately 93.8\% of evaluated cases
using static or combined static--dynamic evidence and it avoids full pipeline processing for the most samples. The static-only route achieves 99.0\%
accuracy, while the static--dynamic route achieves 94.93\% accuracy. Moreover, the static stage
short circuit error rate is limited to 1.14\%. It indicates that early termination can be effective for cases meeting the corresponding routing criteria.

These findings support the suitability of adaptive
evidence acquisition for resource conscious intrusion
detection environments where comprehensive analysis
may introduce unnecessary overhead in terms of cost and delay. Nevertheless, the results also indicate
that adaptive routing must be accompanied by reliable
confidence estimation, particularly for samples
requiring additional analysis.

The cost sensitivity analysis provides further
insight into the operating conditions under which
HMAS is preferable to simpler alternatives.
Under the binary accuracy--cost objective
$\mathrm{Acc}-\lambda C$, HMAS exceeds static-only
analysis for $0 \leq \lambda < 0.01064$
(Table~\ref{tbl:lambda}). This finding indicates
that the preferred analysis strategy depends on
the relative importance assigned to classification
accuracy and computational expenditure.

The binary accuracy--cost objective
does not capture the operational value
of ransomware family attribution or the consequences of missed ransomware detections.
The substantially higher multiclass macro-F1
achieved by HMAS therefore represents an additional
benefit.

The LLM assisted agent introduces a further
trade off between detection sensitivity and resource
consumption. Incorporating LLM review
improves binary recall by 5.3 \%,
although precision decreases by 5.2\%. The reviewer also increases computational
cost and raises decision latency to
22.75~s per case.

In the evaluated configuration, LLM assisted review
reduces family attribution macro-F1 from 0.9748
to 0.9417. Consequently, the deterministic HMAS
configuration currently provides the more favorable
balance for applications prioritizing accurate
family attribution and low latency operation.
It is obvious from results that locally hosted is only benificial for high risks cases where cost may not matter.
From an operational perspective, the proposed framework offers a flexible foundation for
security operations centers (SOCs) and digital forensics and incident response (DFIR) workflows.
The ability to terminate analysis early for
high-confidence cases can reduce processing
requirements and quick action to  patch, while more complex cases can
be escalated for additional inspection.

Overall, the findings establish
adaptive multi-agent orchestration
as a promising approach for
cost-aware ransomware analysis,
particularly for fine-grained
family attribution under
computational constraints.
Addressing the identified
limitations through improved
modality calibration, robust
fusion, selective LLM invocation,
and stronger zero-day evaluation
will further support its
application in practical
intrusion detection environments.
`
\section{Conclusion}
\label{sec:conclusion}

This study presented a cost aware hierarchical multi agent
system (HMAS) for ransomware detection and family attribution,
formulating multimodal evidence acquisition as a budgeted
sequential decision process. Deterministic HMAS achieved 94.12\% binary detection accuracy
(F1 = 0.9376) and, in a separately reported 17-class
evaluation, 98.90\% classification accuracy
(macro-F1 = 0.9748).

Proposed HMAS resolved approximately
93.8\% of cases using static or combined static--dynamic
evidence instead of running full ehxaustive pipeline. This selective processing reduced internal
analysis cost by 44.6\% compared with exhaustive evaluation,
while maintaining strong overall classification performance.
The inclusion of a locally hosted LLM introduced
a performance--efficiency trade off. Although
LLM agent increased binary recall by 5.3\% yet it reduced precision by 5.2 \%, approximately doubled mean analysis cost and
increased decision latency to 22.75~s per case.
These findings suggest that LLM assistance may not be very useful in case of intrusions where time and cost are utmost important.

Overall, the results also demonstrate the potential of
hierarchical multi agent orchestration to balance ransomware
detection, family attribution and time to execute an analysis.

\section{Statements} \label{sect:statements}

\acknowledgement{All work is done on an NVIDIA DGX system. We thank the Pattern Recognition Lab (DCIS), the PIEAS Artificial Intelligence Center (PAIC), and the Center for Mathematical Sciences (CMS), PIEAS for providing computing resources.}

\availabilityofdataandmaterials{The data and supporting materials are available from the corresponding author upon request.}

\ethicsapproval{ This study did not involve human or animal subjects. All malware samples were analyzed in isolated and air gapped virtual machines.}

\conflictsofinterest{The authors declare no conflicts of interest.}

\reftitle{References}

\end{document}